\documentclass[utf8]{frontiersSCNS} 

\usepackage{url,lineno, microtype, subcaption}
\usepackage[onehalfspacing]{setspace}
\usepackage{xcolor}
\usepackage{caption}
\usepackage[finalnew]{trackchanges}
\usepackage{transparent}
\usepackage{colortbl}

\definecolor{bg}{rgb}{.96, .96, .96}

\def\keyFont{\fontsize{8}{11}\helveticabold }
\def\firstAuthorLast{Oliveira {et~al.}} 
\def\Authors{Denny M. Oliveira\,$^{1,2,*}$, 
             Eftyhia Zesta\,$^2$,
             Katherine Garcia-Sage\,$^{2}$}

\def\Address{$^{1}$Goddard Planetary Heliophysics Institute, University of Maryland, 
                   Baltimore County, Baltimore, MD, United States
             $^{2}$NASA Goddard Space Flight Center, Greenbelt, MD, United States
             }

\def\corrAuthor{Denny M. Oliveira}
\def\corrEmail{denny@umbc.edu}

\usepackage[breaklinks, colorlinks = true,
            linkcolor = SkyBlue,
            urlcolor = SkyBlue,
            citecolor = SkyBlue,
            anchorcolor = white,
            breaklinks]{hyperref}

\begin{document}
\onecolumn
\firstpage{1}

\title[Storm-time Starlink reentry]{Tracking Reentries of Starlink Satellites During the Rising Phase of Solar Cycle 25} 

\author[\firstAuthorLast ]{\Authors} 
\address{} 
\correspondance{} 

\extraAuth{}

\maketitle

\begin{abstract}

    The exponential increase of low-Earth orbit (LEO) satellites in the past 5 years has brought into intense focus the need for reliable monitoring and reentry prediction to safeguard from space collisions and ground debris impacts. However, LEO satellites fly within the upper atmosphere region that exerts significant drag forces to their orbits, reducing their lifetimes, and increasing collision risks \add{during dynamic events, like geomagnetic storms}. Such conditions can become more severe during geomagnetic storms, particularly during extreme events. In this work, we use two-line element (TLE) satellite tracking data to investigate geomagnetic activity effects on the reentries of 523 Starlink satellites from 2020 to 2024. This period coincides with the rising phase of solar cycle 25, which has shown itself to be more intense than the previous solar cycle. We derive satellite altitudes and velocities from TLE files and perform a superposed epoch analysis, the first with hundreds of similar satellites. Even with limitedly accurate TLE data, our results indisputably show that satellites reenter faster with higher geomagnetic activity. This is explained by the fastest orbital decay rates (in km/day) of the satellites caused by increased drag forces. We also find that prediction errors, defined as the difference between the epochs of actual reentries and predicted reentries \add{at reference altitudes}, increase with geomagnetic activity. As a result, we clearly show that the intense solar activity of the current solar cycle has already had significant impacts on Starlink reentries. This is a very exciting time in satellite orbital drag research, since the number of satellites in LEO and solar activity are the highest ever observed in human history.



    \tiny
    \keyFont{ \section{Solar activity, geomagnetic storms, thermosphere response to storms, satellite mega-constellations, satellite orbital drag, satellite reentry}} 
\end{abstract}

\section{Introduction}

    From the launch of Sputnik 1 in 1957, the first artificial satellite by the Soviet Union, the history of satellites in space has been marked by rapid technological advancements and significant geopolitical impacts. Sputnik's success triggered the space race between the United States and the Soviet Union, culminating in the creation of the first communications satellites, weather satellites, and reconnaissance satellites throughout the Cold War \citep{Cruddas2019}. In the decades that followed, satellite technology evolved to support global communication, scientific research, and the launch of GPS satellites revolutionized navigation. In the 21st century, private companies like SpaceX, with its Starlink mega-constellation, have redefined the role of satellites by developing massive, low-Earth orbit (LEO) networks to provide global internet access \citep{Ren2022}. This era represents a shift from government-driven space exploration to commercial innovation, illustrating how satellites have transitioned from a strategic military tool to an integral part of everyday life, connecting the world in ways once imagined only in science fiction \citep{Moltz2019}. \par

    The Kessler Syndrome, suggested by \cite{Kessler1978}, describes a self-perpetuating cascade of satellite collisions in LEO, where the impact of one satellite or piece of debris generates additional fragments. These fragments, traveling at high speeds, can collide with other objects, creating even more debris. If this cycle continues, it will exponentially increase the number of debris in orbit, making space more hazardous for satellites and spacecraft. The critical threshold is reached when the density of debris becomes high enough that collisions occur frequently, leading to a dangerous feedback loop. As a result, the Kessler Syndrome could make certain regions of space, particularly LEO, increasingly unusable for future satellites and space missions, severely impacting operational, exploratory, and commercial activities \citep{Boley2021,Witze2022}. Therefore, being able to predict such collisions to avoid further debris generation in LEO is a very important space weather research topic with significant societal impacts \citep{Bruinsma2021,Parker2024}, particularly during extreme geomagnetic conditions \citep{Oliveira2021a}. \par

    Geomagnetic storms, which are disturbances in Earth's magnetic field caused by solar wind perturbations like coronal mass ejections (CMEs), can significantly heat the upper atmosphere, particularly in LEO \citep{Fuller-Rowell1994,Prolss2011,Emmert2015}. When charged particles from the solar wind interact with Earth's magnetosphere, they intensify electric currents in the ionosphere which enhance levels of collisions between ionospheric plasma and thermospheric neutral atoms and molecules. The additional kinetic energy is locally dissipated as Joule heating \citep{Fedrizzi2012,KalafatogluEyiguler2019a}, making the thermosphere become denser and upwell to altitudes where satellites orbit \citep{Liu2005b,Prolss2011}. Solar radiation affects thermosphere neutral mass density by increasing temperature and causing atmospheric expansion, which leads to higher density at a given altitude \citep{Emmert2015}. During periods of high solar activity, enhanced extreme ultraviolet (EUV) and X-ray radiation increase ionization and heating, further amplifying density variations \citep{Qian2011,Emmert2015}. Thermospheric neutral mass density is directly associated with drag acceleration, which is naturally linked to increased drag forces on satellites \citep{Sutton2005,Calabia2017,Bruinsma2023}. This heightened atmospheric density results in greater orbital drag on satellites, as they experience increased resistance from the denser and hotter thermosphere \citep{Prieto2014,Oliveira2019b,Krauss2020}. Geomagnetic storms lead to variations in the satellite’s orbit, as the increased atmospheric drag can change the satellite's velocity and altitude over time. This effect is particularly pronounced for satellites in very-low Earth orbits (VLEO, usually considered to be bellow 300 km), where the atmosphere is denser, and the drag force is more significant. Consequently, geomagnetic storms can shorten a satellite's operational lifespan, increase fuel consumption for orbital maintenance, and complicate mission planning and reentering. \par

    In February 2022, nearly 40 Starlink satellites reentered Earth's atmosphere before reaching their intended operational altitude (500-600 km) due to a combination of a few factors. First, the satellites were deployed to a low altitude (210 km) before thrusters were expected to boost up the satellites \citep{Hapgood2022}. Second, Starlink CONOPs (Concepts of Operations) required that the satellites' solar panels kept pointing toward the Sun \citep{Mallama2024}, thus increasing drag. Third, the satellites were deployed while a minor geomagnetic storm was raging on. Consequently, that storm contributed to the increase of the local neutral mass density, which subsequently amplified drag forces even further \citep{Fang2022,Berger2023,Baruah2024}. As demonstrated by \cite{Fang2022}, the thermosphere density underwent an increase of nearly 50\% in comparison to density values before the storm. For these reasons, the satellites decayed faster than anticipated. As a result, the satellites were prevented from using their fuel to raise their orbits as planned, resulting in their loss before they could reach their operational altitudes. \change{Thus, accurately tracking Starlink reentries is of paramount importance to avoid collisions and safeguard the ground from potential debris.}{Thus, having better reentry predictions can greatly improve operational decisions at VLEO and aid in the planning of controlled reentries of artificial space objects.} \par


    In this work, we follow on the work provided by \cite{Oliveira2025a}, who performed a case study using orbital tracking data of four Starlink satellites reentering during different geomagnetic conditions. The authors noted that, the more intense the geomagnetic condition, the faster the satellite orbital decay rate (in km/day). \cite{Oliveira2025a} also noted that reentry prediction errors were higher during more intense geomagnetic conditions. Here, we use similar Starlink orbital data to perform a superposed epoch analysis of orbital altitudes and velocities in order to identify impacts caused by storms with different intensities. The Starlink reentries coincide with the rising phase of solar cycle (SC) 25, a period with increasing solar activity. By confirming the case study of \cite{Oliveira2025a}, we highlight the importance of accurately predicting and tracking satellite reentries during storm times, particularly during severe geomagnetic storms. Improving such capabilities is of paramount importance to avoid the triggering of the Kessler syndrome which could make satellite traffic in LEO nearly inoperable. This study, with hundreds of similar satellites, is the first of its kind. Our work points the way for determining continuous prediction of reentry time and place based on geomagnetic conditions.

 \section{Materials and methods}

    \subsection{Solar activity}

        Solar activity is represented by the F10.7 solar radiation index \citep{Tapping2013}. F10.7 represents the flux density of solar radio emissions at a wavelength of 10.7 cm, measured in solar flux units (sfu), where 1 sfu = 10$^{-22}$ W$\cdot$m$^{-2}\cdot$Hz$^{-1}$. It serves as a key indicator of solar activity and is strongly correlated with other solar parameters such as sunspot numbers and extreme ultraviolet (EUV) emissions \citep{Hathaway2015}, which influence Earth's upper atmosphere \citep{Qian2011,Emmert2015}. The F10.7 index is measured daily by ground-based radio telescopes, primarily at the Dominion Radio Astrophysical Observatory in Canada. It is widely used in space weather research, as it helps estimate thermospheric density, ionospheric conditions, and satellite drag. Unlike sunspot numbers, which are subjective visual counts, F10.7 provides an objective and continuous measure of solar variability, making it a reliable proxy for solar activity over both short-term (solar rotation) and long-term (solar cycle) timescales.

    \vspace{0.6cm}
    \subsection{Geomagnetic activity}

        Geomagnetic activity is represented by the storm-time, 1-hour Dst index \citep{WDC_Dst2015}. According to \cite{WDC_Dst2015}, the real-time Dst index provides quick, preliminary values of geomagnetic storm activity with minimal data processing, making it slightly less accurate than its final version. The provisional Dst index is more refined, with some error corrections applied, but is still subject to further adjustments. The final Dst index is the most accurate, as it undergoes extensive quality control, calibration, and correction, providing the most reliable data for scientific analysis after a storm. The time span of our analysis covers all versions of Dst data: final (2020), provisional (2021-2023), and real-time/quick-look (2024).

    \vspace{0.6cm}
    \subsection{Satellite orbital parameters}

        Satellite orbital information is extracted from Two-Line Element (TLE) data files. TLE data files are a standard format used to represent the orbital parameters of artificial satellites and other objects in space. Each TLE consists of two lines of data, with the first line containing information about the satellite’s identification, classification, and epoch (the specific time at which the orbital parameters were valid), while the second line includes key orbital elements like inclination, eccentricity, altitude, and the satellite's position and velocity in its orbit \citep{Kizner2005}. TLEs are updated regularly, usually every few days, to reflect changes in the satellite's orbit due to gravitational forces, atmospheric drag, or maneuvers. These data files are widely used to track satellites, predict passes, and monitor space debris. The format is compact and machine-readable, making it ideal for quick computation of satellite positions and orbits. \par

        TLE data is downloaded from \texttt{space-track.org}, a website that provides free access to satellite data. The platform is maintained by the U.S. Department of Defense and offers real-time tracking data, including information on active satellites, debris, reentry predictions, and other objects in Earth’s orbit. Access requires creating a free account. \par

        The Python package \texttt{PyEphem} is used to determine positions (altitudes, latitudes, and longitudes) and velocities from TLE data. \texttt{PyEphem} provides a simple and efficient way to extract orbital information from TLE data files, allowing users to compute satellite positions and track objects in orbit. \texttt{PyEphem} uses the TLE format to initialize a satellite object, parsing the two lines of data into orbital parameters such as inclination, eccentricity, right ascension of the ascending node, and perigee. The package then applies the SGP4 (Simplified General Perturbation Model 4) algorithm to propagate the satellite’s orbit over time, accounting for various gravitational forces and orbital perturbations \citep{Acciarini2025}. Once the TLE data is loaded and processed, \texttt{PyEphem} can provide the satellite's position (in terms of azimuth, elevation, or geodetic coordinates) for any given time, enabling predictions of satellite passes or orbital changes. This makes \texttt{PyEphem} an invaluable tool for satellite tracking, astronomy, and space-related research. \par

        In a TLE file, the mean motion $(n)$ of a satellite represents the number of orbits that the satellite completes per day. Thus, the mean motion can be used to compute the velocity of a satellite at a given epoch according to the formula \citep{Curtis2005}
            \begin{equation}
                v = (GM\eta)^{1/3}\,,
            \end{equation}
        where $G$ is the universal gravitational constant, $M$, the Earth's mass, and $\eta=2\pi n/T$, with $T$ = 86400 s being the number of seconds in a day. \par

        In this paper, we consider a satellite (or its remaining debris) reenters from VLEO when it crosses the K\'arm\'an line at 100 km altitude, commonly considered an altitude threshold from satellites reentering from outer space into the terrestrial atmosphere \citep{Karman1956,McDowell2018}. However, in most cases, the altitude of the last epoch of a given satellite is in the interval 180-140 km. For this reason, we use \texttt{PyEphem} to propagate the satellite's altitude in time until it crosses the K\'arm\'an line. \add{Most of the altitudes are propagated in time intervals within 1 day from the last TLE epoch.} Then, we use Newtonian mechanics to estimate the satellites’ velocity according to the expression:
            \begin{equation}
                v = \sqrt{\frac{GM}{R(\varphi) + h}}\,
            \end{equation}
        with $h\sim100$ km, and $R(\varphi)$ being the geodetic Earth's radius at the specific ground latitude $\varphi$ \citep{Torge1980}. Circular orbits are assumed in the use of the previous equations. \par

        \subsection{Superposed epoch analysis parameters}

            {\bf Zero-epoch time:} a zero-epoch time for each reentry is chosen when the satellite commences its sharper decay. This reference altitude is taken as close to 280 km altitude as possible, and it is usually between 260 km and 320 km. The zero-epoch altitude is based on the inspection of altitudes during Starlink reentries under intense and severe geomagnetic conditions. An example will be shown later. Figure 2 of \cite{Oliveira2025a} also shows a few more examples. \par

            \add{\protect{{\bf Tracking and Impact Prediction (TIP) messages:} a TIP message provides epoch information of the most accurate reentry epoch. According to \texttt{space-track.org}, TIP messages begin to be issued a few days prior to the corresponding object's decay. Such messages are also issued multiple times in the last 24 hours preceding the decay. Although TIP messages bring latitude and longitude information, they do not provide altitude information.}} \par

            {\bf Reentry prediction errors:} a reentry prediction epoch is obtained for each satellite. We use reentry predictions provided by TIP messages for the day of the zero-epoch time. If space-track.org does not provide a reentry prediction for the reference altitude epoch, we apply the SGP4 propagator to the TLE data corresponding to the zero-epoch time to propagate the satellite's orbit until it crosses the Kármán line. \remove{Although the Tracking and Impact Prediction (TIP) messages provided by space-track.org are accurate, we use reentry predictions at the reference altitude epoch for comparisons with actual reentries to account for solar and geomagnetic activity effects on the satellites' reentries. As a result, the reentry prediction error is defined as the time difference, in days, between the epochs of reference altitude and the actual reentry.}  \par

            \add{\protect{{\bf Estimated reentry:} this is the estimated epoch obtained by propagating the last TLE data file until the altitude reaches 100 km. We use this approach to obtain our estimated reentries as opposed to using the latest available TIP messages because the latter are issued when satellites are at many different altitudes above 100 km. As a result, our approach allows for more adequate reentry altitude estimates in the superposed epoch analysis.}} \par

            {\bf Day difference:} Day difference is simply the time interval between the \change{actual}{estimated} reentry epoch and the reference altitude (zero) epoch represented in days. \par

            \begin{figure}
                \centering
                \includegraphics[width = 16cm]{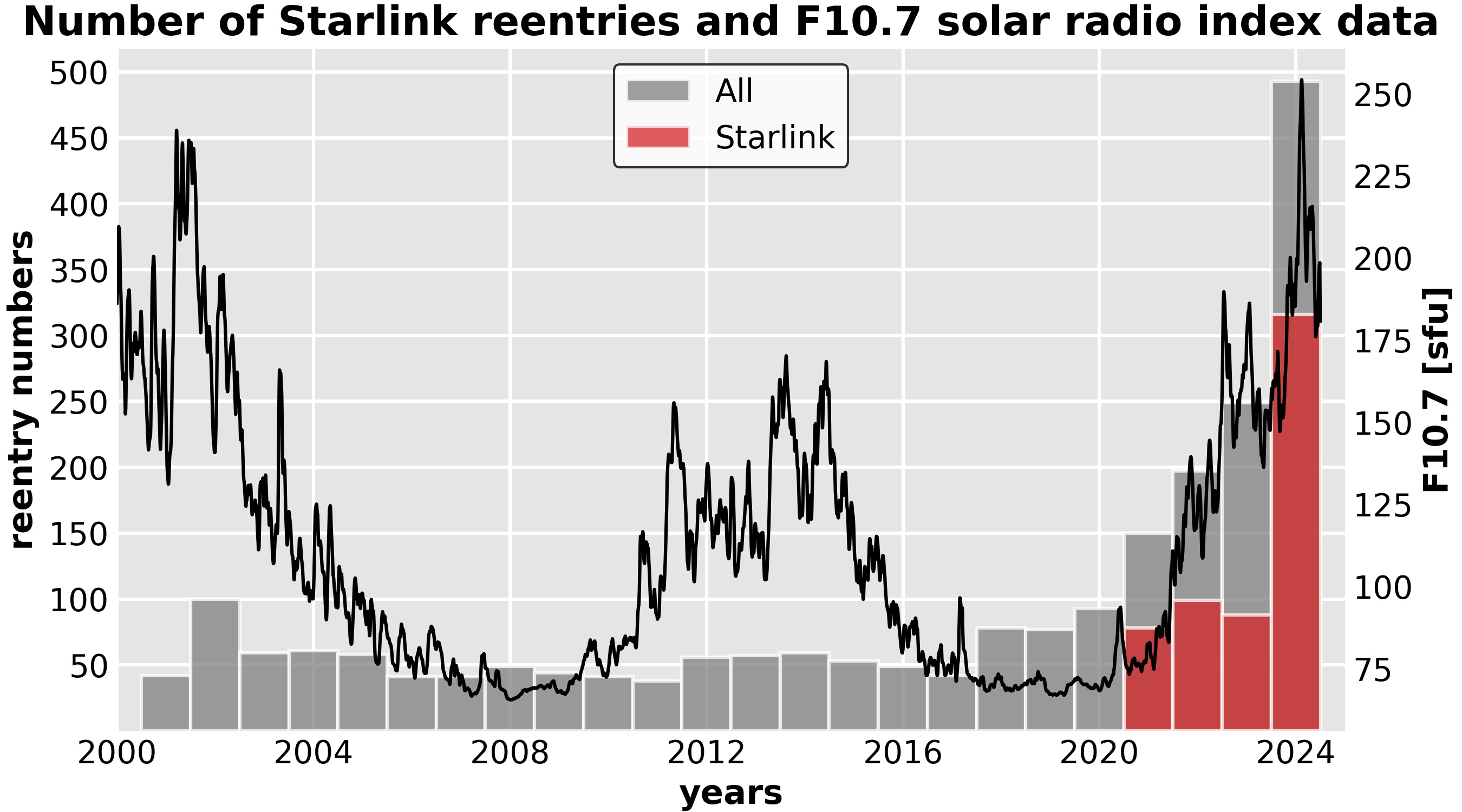}
                \caption{Annual number distribution of Starlink reentries and monthly-averaged F10.7 solar radio flux index data from 2000 to 2024. Reentry data was obtained from \texttt{space-track.org}. The grey bars indicate all satellites, and the red bars indicate Starlink satellites.}
                \label{reentries}
            \end{figure}

            It is important to emphasize that the use of TLE data and the SGP4 model in their analyses is considered unsuitable for accurate reentry predictions, except for providing the starting conditions for numerical integration if more accurate ephemeris data is unavailable. SGP4 does not include any atmospheric model, but it just extrapolates the observed changes in mean motion, and can therefore not account for rapid increases in atmospheric densities at lower altitudes in the days before reentry. However, though limited, SGP4 can provide a prediction baseline for indirect comparison between orbital drag effects on Starlink satellites reentering during different solar and geomagneic conditions. This is the main goal of this work.

    \section{Statistical results and discussion}

        \subsection{Reentries}

            Figure \ref{reentries} shows annually number distributions of satellites (colored bars) and monthly-averaged solar radio flux index data (black line) for the period of 2000-2024. Grey bars indicate data for all satellites, whereas red bars indicate data for Starlink satellites. \par

            The data plotted in the figure covers a time interval equivalent to two solar cycles. This includes the declining phase of SC23, the entire SC24, and the rising phase of SC25. The plot clearly shows that before 2021, the maximum number of reentries was near 100, with most years presenting reentry numbers around 50. In 2019, SpaceX started launching Starlink satellites into LEO and they began reentering in late 2020. Starlink reentry numbers were relatively low in the first three years, with 2 reentries in 2020 (not visible in the histogram), 78 in 2021, 99 in 2022, 88 in 2023, and an impressive number of 316 reentries in 2024. The intermediary number between 2021 and 2023 may have been impacted by the unexpected reentries of 39 Starlink satellites in February 2022 \citep{Hapgood2022}. From 2020 to 2024, 1190 satellites reentered from VLEO, with 583 (nearly half) being Starlink satellites. \par

            \cite{Oliveira2021a} pointed out that the number of satellites in LEO would increase dramatically in the following years after their publication. Additionally, they mentioned that those numbers would coincide with increased solar activity in SC25. This can be seen in the rising phase of the current solar cycle (2020-2024). Therefore, it is clear that SpaceX has made a significant impact on the number of reentries of satellites from VLEO. As a result, such reentries ought to be closely tracked to avoid collisions with other satellites or debris, as well as downfall of Starlink debris in undesirable regions, such as populated areas. Such tracking should be performed even more cautiously during periods of high solar activity. \par

    \begin{figure}
        \centering
        \includegraphics[width = 17cm]{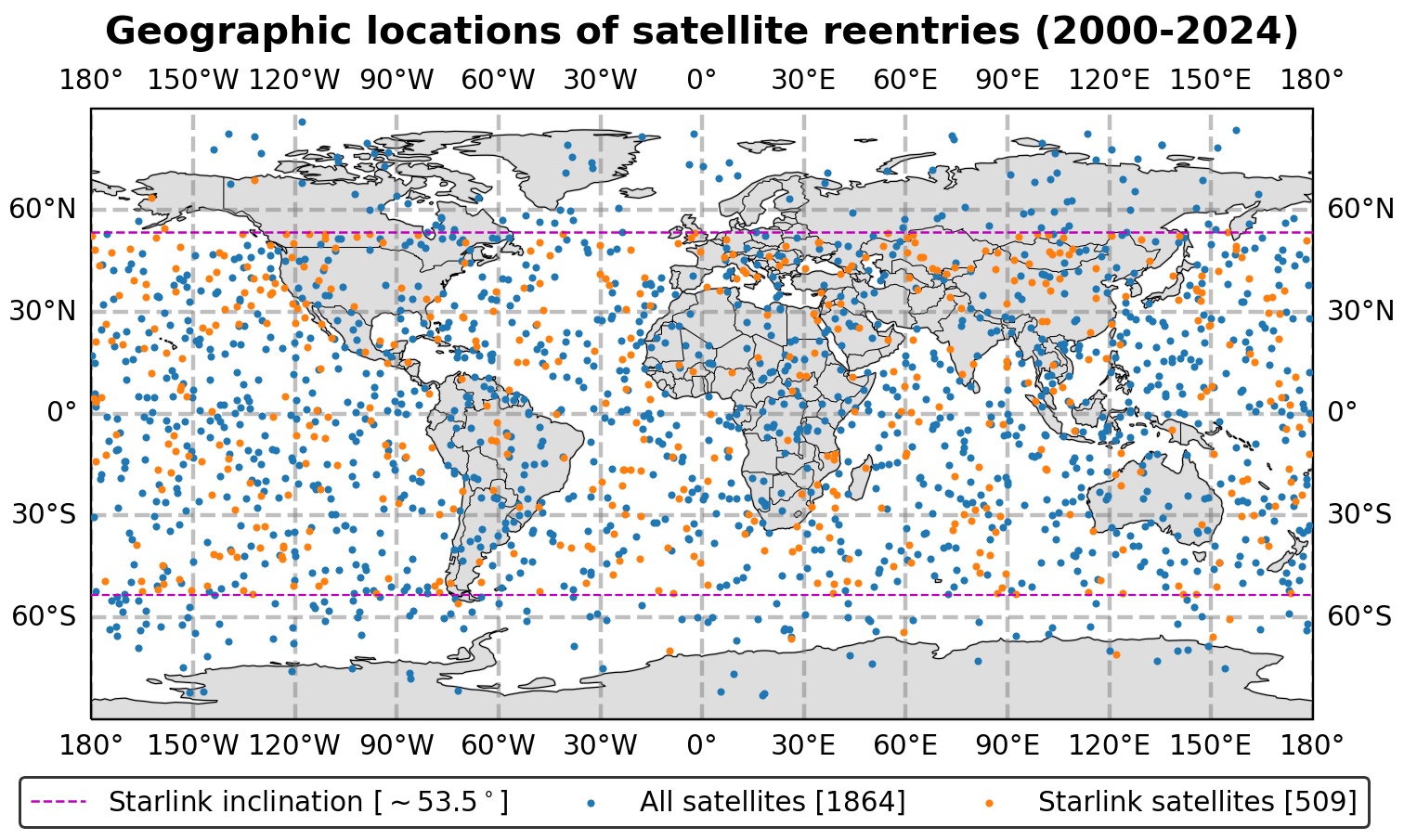}
        \caption{Geographic reentry locations of 1864 satellites in the period 2000 to 2024 \add{obtained from available TIP message data}. Orange dots are for Starlink satellites (509), and blue dots are for non-Starlink satellites (1355). Both magenta dashed horizontal lines show the orbit constrained region by Starlink inclinations  ($\sim$ 53.5$^\circ$).}
        \label{starlink_world}
    \end{figure}

            The world map shown in Figure \ref{starlink_world} displays reentry geographic locations of all the \change{2233}{1864} satellites \add{with available TIP message data} found at \texttt{space-track.org} within the 2000-2024 period. Orange dots indicate Starlink satellite (509) positions, whereas blue dots indicate the positions of the remaining 1355 satellites. The horizontal dashed magenta lines delimit typical orbital latitudes due to the inclinations of the Starlink satellites ($\sim\pm53^\circ$). As can be seen, Starlink satellites reentered in most regions of the globe within the inclination limits, particularly over the ocean. \change{However, there were slightly more reentries over a relatively highly populated area in central Europe. On the other hand, the other satellites reentered in the following concentrated regions: east and west coasts of the United States, the Caribbean, Europe, low latitudes in the Pacific Ocean, northeast of Australia, and along the equatorial line. While we are not able to specify through TLEs the precise location of atmospheric burn and potential impact to the ground of any remaining debris, with so many reentries over human populations and assets,}{Although fewer satellites reentered over land,} it is important to track reentries of satellites from VLEO to safeguard the ground from their reentries according to international norms, including debris mitigation \citep{Ailor2007}. \par    

    \begin{figure}[t]
        \centering
            \includegraphics[width = 11cm]{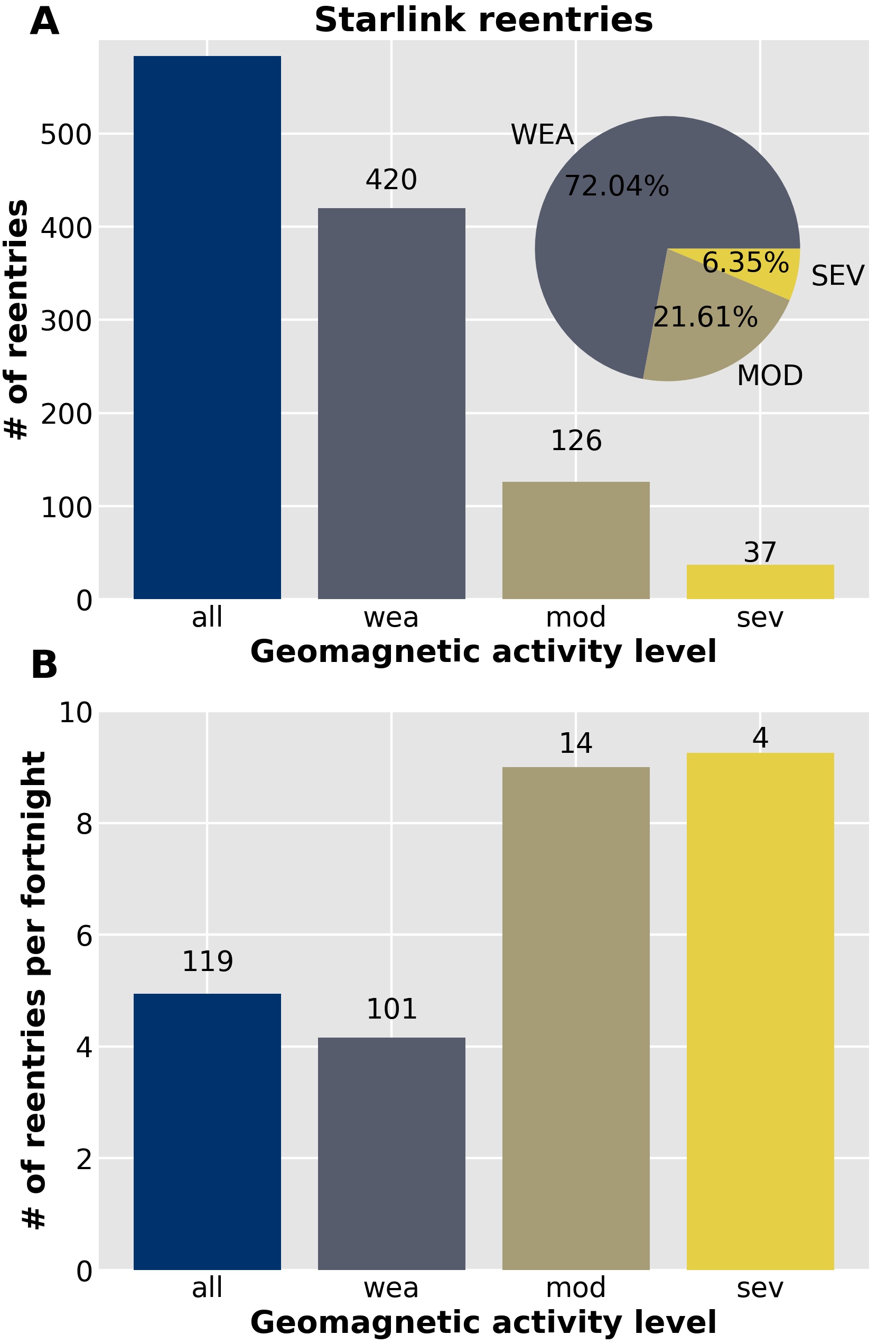}
            \caption{{\bf A:} Statistical number distributions of all events and events grouped in the three different geomagnetic activity levels (weak, moderate, and severe). The inset pie diagram shows the relative distributions of the reentries under the three different geomagnetic conditions here investigated. {\bf B}: Number of Starlink reentries per fortnight (14 days) with minimum Dst values falling in the storm level definitions presented in section 3.1. \add{The numbers at the top of the bars represent the numbers of fortnights within each geomagnetic level category.}}
            \label{storm_levels}
    \end{figure}

            In this paper, we group reentries occurring during three different geomagnetic activity levels: weak, moderate, and severe. Such groups are arbitrarily defined as follows:

                \begin{center}
                    \begin{tabular}{l c}
                        Weak: \quad & Dst $>$ --100 nT \\
                        Moderate: \quad & --200 nT $<$ Dst $\leq$ --100 nT \\
                        Severe: \quad & Dst $\leq$ --200 nT
                    \end{tabular}
                \end{center}

            Figure \ref{storm_levels}A shows statistical number distributions of all Starlink reentries occurring during three geomagnetic conditions. The number of events within each geomagnetic activity category, represented at the top of each vertical bar, decreases as the geomagnetic activity becomes more intense. The inset pie diagram shows the relative statistical distributions of each reentry storm category. Despite the relatively low number of reentries during severe geomagnetic condition (37 reentries or 6.35\% of all reentries), we will show that severe geomagnetic storms have already significantly impacted the de-orbiting of Starlink satellites during the rising phase of SC25. \par

            Since the number of reentries occurring during moderate and severe geomagnetic conditions are relatively low, we also determine the number of fortnight (14 day) intervals with minimum Dst values falling within each storm activity level interval. We then estimate the number of Starlink reentries per fortnight, and the results are shown in Figure \ref{storm_levels}B. The numbers at the top of the bars indicate the numbers of fortnights occurring during each corresponding geomagnetic activity level. Results show that there were $\sim$5 reentries per fortnight during the whole period (119 fortnights), $\sim$4 reentries per fortnight during weak geomagnetic conditions (101 fortnights), around 9 reentries per fortnight during moderate (14 fortnights) and severe (4 fortnights) geomagnetic conditions. Overall, Figure \ref{storm_levels} clearly shows that the number of reentries decreases with solar activity being particularly low during severe conditions, but the reentry rates (reentries per fortnight) during moderate and severe geomagnetic conditions are nearly identical ($\sim$9 reentries per fortnight).\par

            It is important to note that we only have real-time/quick-look or provisional Dst data available, so that is what we used for the reentries displayed in Figure \ref{storm_levels}. Although the change of non-final version data to their final version may affect the weak and moderate categories, as argued by \cite{Oliveira2025a}, it is very unlikely a severe event could be lowered to a moderate event. Nevertheless, such changes would most likely be unnoticeable given the large number of satellite reentries here investigated.

        \vspace{0.6cm}
        \subsection{Superposed epoch analysis}  

            We use the Gannon superstorm of May 2024 as an example to illustrate our methodology. The Gannon superstorm was the most intense event in two decades since the Halloween events of 2003 \citep{Hayakawa2025}. \cite{Hayakawa2025} tracked the sunspot group and X-class solar flares associated with the very fast CMEs that drove the storm. The authors concluded that a large amount of magnetic energy was transferred into the magnetosphere leading to a very extreme geomagnetic storm. As a result, many accounts of low-latitude auroras were recorded around the world \citep{Gonzalez-Esparza2024,Hayakawa2025}. \par

                \begin{figure}[t]
                    \centering
                    \includegraphics[width = 14cm]{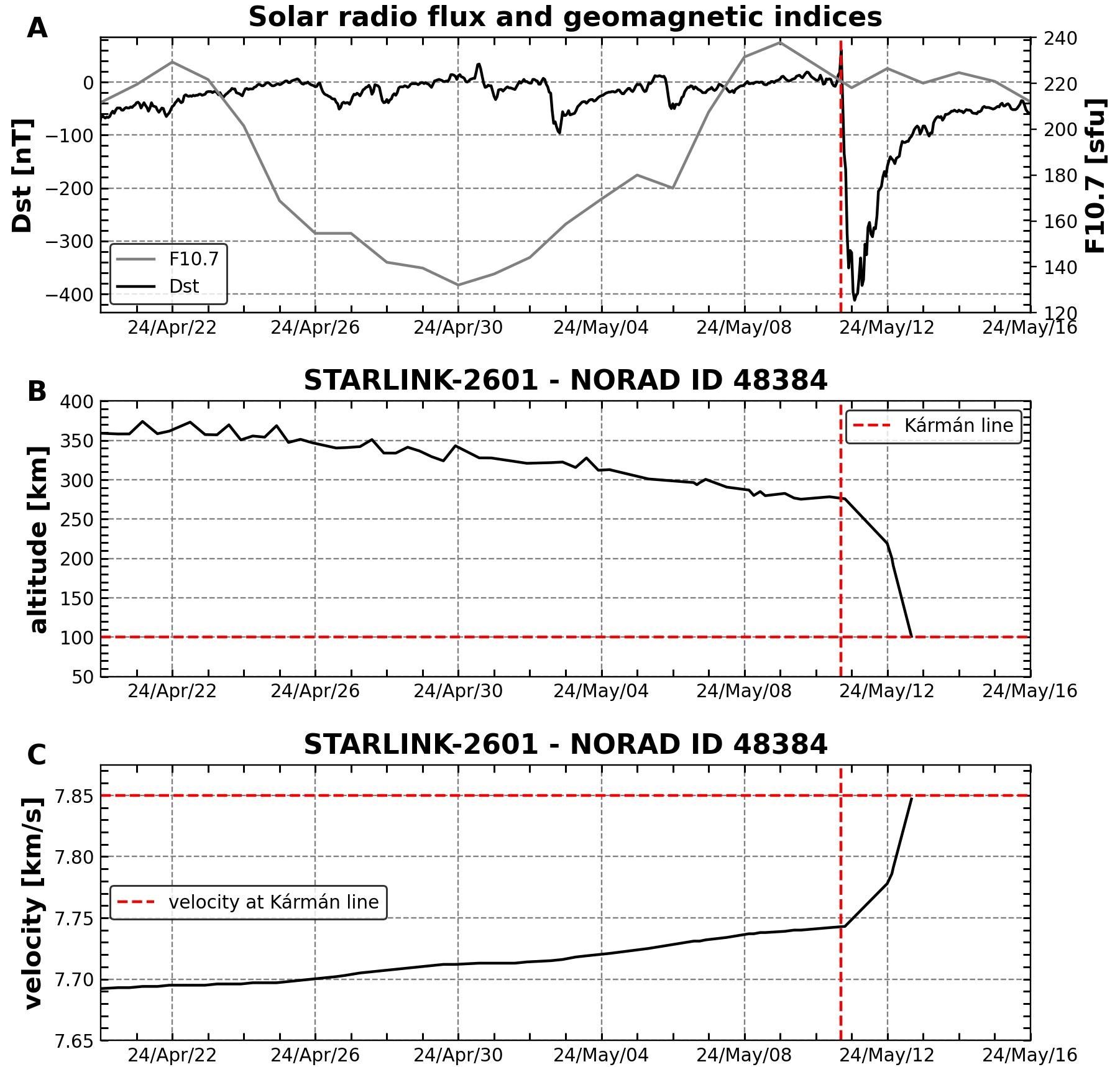}
                    \caption{An example of how solar and geomagnetic activity can impact a satellite reentering from LEO. The figure shows the Starlink-2601 satellite (NORAD ID 48384) reentering during the May 2024 Gannon storm. {\bf (A)}: hourly values of Dst index (solid black line) and daily F10.7 solar radio flux index (solid grey line); {\bf (B):} nearly daily satellite altitudes; and {\bf (C)}: the same for velocities. The red dashed vertical lines in all frames indicate the storm sudden commencement at $\sim$ 1700 UT on 10 May 2024, which nearly coincides with the reference altitude epoch at 1930 UT on the same day. The red dashed horizontal lines in frames {\bf (B)} and {\bf (C)} indicate the K\'arm\'an line and velocity at 100 km altitude, respectively.}
                     \label{starlink_may2024}
                \end{figure}

            Dst data for that storm is plotted in Figure \ref{starlink_may2024}A. Its minimum Dst value is --412 nT, which followed the storm sudden commencement (SSC) at nearly 1700 UT \citep{Piersanti2025}. The frame also shows that F10.7 increases within 8 days from $\sim$130 to $\sim$ 240 sfu before storm onset, and stays close to 220 sfu during the storm period. Frame B shows TLE altitudes for the Starlink-2601 (NORAD ID 48384) satellite. The spacecraft was already being decommissioned when the SSC took place \citep[see discussion in][]{Oliveira2025a}. More coincidently, Starlink-2601 crosses the reference altitude (dashed vertical red line) of 276 km at 1930 UT on May 10. Consequently, the satellite plummets down to the K\'arm\'an line at 100 km in a time interval of 1.86 days, resulting in an altitude drop of 176 km at the orbital decay rate of 95 km/day. As expected, the satellite velocity increases due to the conversion of its gravitational potential energy into kinetic energy. The velocity at the K\'arm\'an line, indicated by the horizontal dashed red line, is about 7.85 km/s (Figure \ref{starlink_may2024}C). We propagate TLE data for the reference altitude epoch to obtain a reentry prediction epoch at 0000 UT of 23 May 2024, but propagation of the last TLE record indicates that the satellite reached the K\'arm\'an line at 1609 UT of 12 May 2024, $\sim$11 days before prediction. These observations agree with the comparative case study provided by \cite{Oliveira2025a}. The combination of high solar and geomagnetic conditions after 10 May 2024 contributed to the fast decay of Starlink-2601 shown in Figure \ref{starlink_may2024}B. \par

            \add{For the superposed epoch analysis, we selected altitudes near 280 km as the zero epoch time. Even for extreme storms such altitudes would provide sufficient opportunities for a satellite to elevate and make altitude or re-entry decisions. TIP messages are only issued at very low altitudes, below 200 km, when it is too late for commanding decisions. It is important to note that this is the altitude that has emerged as the ``last change" altitude for the Starlink satellite. It is possible and even likely that for a different satellite configuration (volume, shape, mass) such a critical altitude could be different.} \par

            Starlink reentries during the Gannon storm are included in our superposed epoch analysis as well. We also include Starlink reentries during the extreme storm of October 2024, the second most extreme event since October 2003. \cite{Xia2025} showed that field-aligned currents during that storm were quite strong and reached low-latitude regions, along with distinct visible auroras. \cite{Oliveira2025a} also showed a Starlink satellite reentered sooner than expected in October 2024. \par

\begin{figure}[t]
                    \centering
                    \includegraphics[width = 14cm]{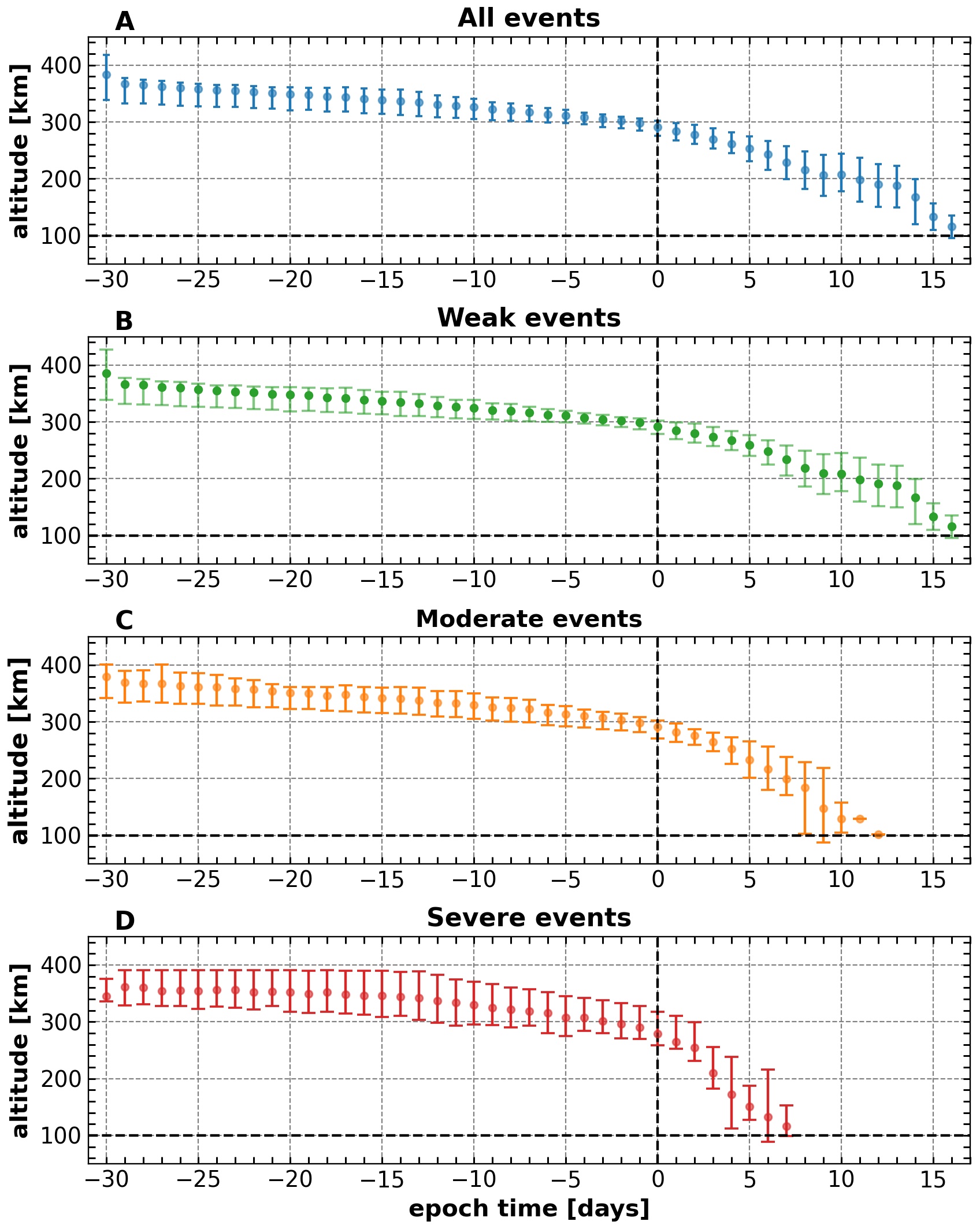}
                    \caption{Superposed epoch analysis of 523 Starlink altitudes for reentries during different levels of geomagnetic activity. Frame A, all events; B, weak events; C, moderate events, and D, severe events. The error bars indicate 25th percentiles (lower edge) and 75th percentiles (upper edge). The black dashed vertical lines indicate the zero epoch time taken as the reference altitude epoch for each satellite. The black horizontal line represents the K\'arm\'an line (altitude 100 km).}
                    \label{xht_sea}
                \end{figure}

            Figure \ref{xht_sea} shows results of superposed epoch analysis for 523 Starlink satellites (TLEs of 60 satellites were either unavailable or incomplete). Data is superposed in a time interval of 30 days before and 16 days after the zero epoch time. In all panels, the solid dots indicate mean altitude values at a daily cadence, whereas the error bars represent the 25th (lower edge) and the 75th (upper edge) quartiles of the distribution for the epoch day. In the plot, frame A is for all events, B is for weak events; C, moderate events; and D, severe events. \par

            As can be seen from the figure, all satellites were in reentry processes during the time period here investigated. In the case of all events (A), altitude error bars were relatively small around the zero epoch time (a few kilometers). The average reentry time for these satellites is 16 days. However, upon reentries ($t$ $>$ 0), error bars significantly increase, particularly after day 5 of reentry. \par

            Results shown for reentries during weak conditions are quite similar to results shown for all reentries, given that reentries during weak conditions correspond to $\sim$ 72\% of all reentries (Figure \ref{xht_sea}B). Figure \ref{xht_sea}C shows that the average reentry times during moderate geomagnetic conditions is 12 days from the time the satellite reached the reference altitude. The frame also shows that reentry altitudes during moderate geomagnetic conditions show larger error bars, reaching more than 100 km on day 9 of reentry. For severe conditions (Figure \ref{xht_sea}D), error bars are quite large during the whole time period, being near 100 km most of the time. Average reentry times during severe geomagnetic conditions, 7 days, are the shortest. \par

            The large altitude error bars for moderate and severe geomagnetic conditions are presumably related to the relatively low number of reentries during moderate geomagnetic conditions (126), and particular of satellites reentering during severe geomagnetic conditions. Those latter conditions correspond to only 37 reentries ($\sim$ 6\%) since we are considering here only the rising phase of SC25. In addition to its low numbers, the severe geomagnetic storm group has the largest Dst variability (--200 nT to --412 nT) which can also explain the large error bars for this category. Satellite maneuvers, including possible continuous thrusting before and orbit lowering during the reentries, may also play a role in producing large altitude error bars since some maneuvers may occur to avoid collisions with debris or optimize satellites' orientations during reentries. Despite this relatively short period and low reentry numbers during elevated geomagnetic activity levels, our results clearly show that solar activity has already had significant impacts on Starlink reentries. Our results confirm the comparative case study results of \cite{Oliveira2025a}. \par

                \begin{figure}[t]
                    \centering
                    \includegraphics[width = 14cm]{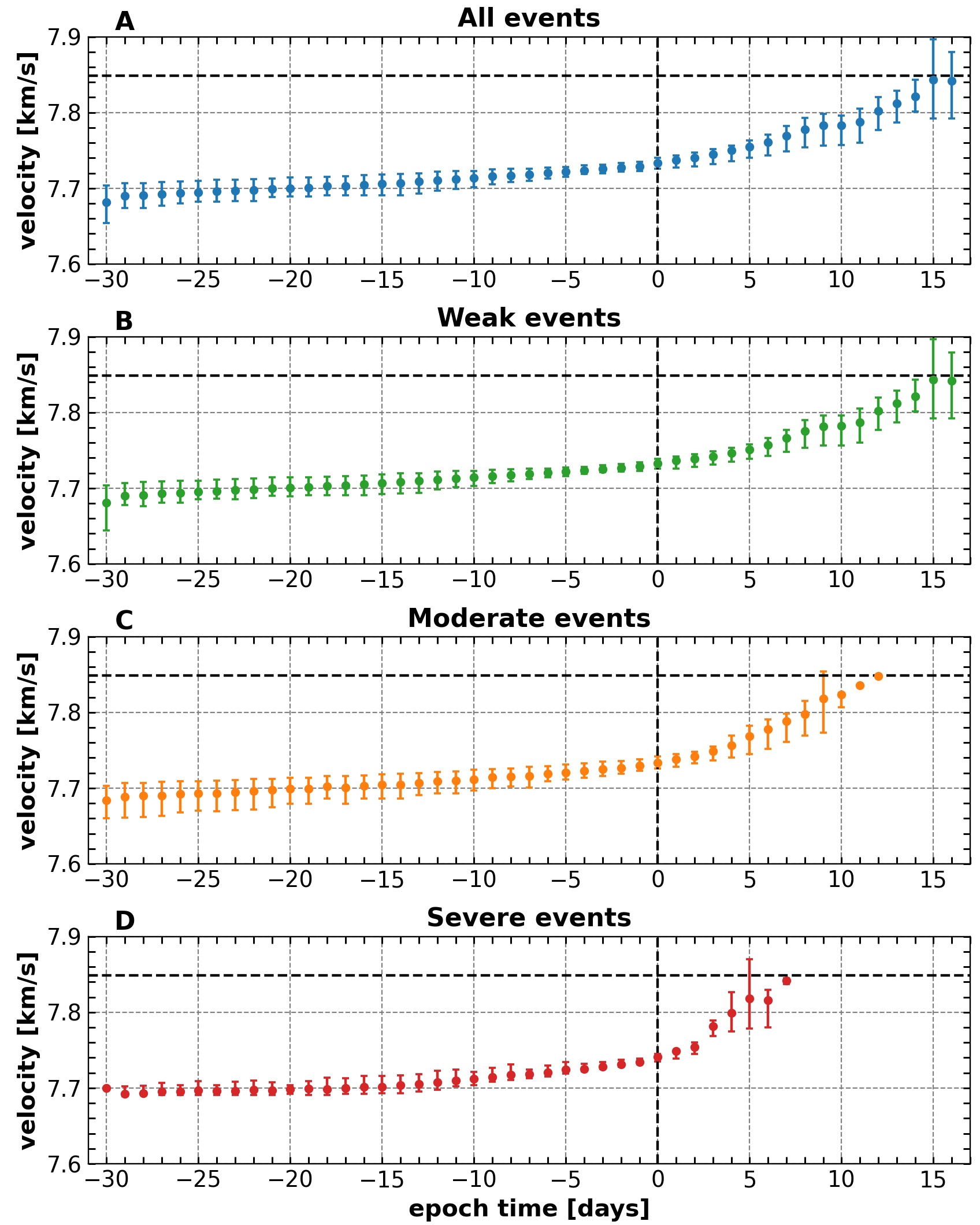}
                    \caption{The same as in Figure \ref{xht_sea}, but for Starlink velocities in km/s. The black dashed horizontal line indicates the K\'arm\'an velocity ($\sim$ 7.85 km/s).}
                    \label{vel_sea}
                \end{figure}

            Figure \ref{vel_sea} for velocities shows similar results as seen in Figure \ref{xht_sea} for altitudes during reentries under different geomagnetic conditions. Here, we clearly see in the figure that satellites reach the K\'arm\'am velocity ($\sim$ 7.85 km/s) faster for increasing geomagnetic conditions, even using limited daily TLE data and simplified circular orbit assumptions. This is a clear consequence of how fast gravitational potential energy becomes kinetic energy as a function of geomagnetic activity. Such results confirm the case example seen in Figure \ref{starlink_may2024} and the case study shown in figure 2 of \cite{Oliveira2025a}. Therefore, this velocity analysis is here included simply as a complement to the results discussed in the previous figure. \par

\begin{figure}[t]
                    \centering
                        \includegraphics[width = 11cm]{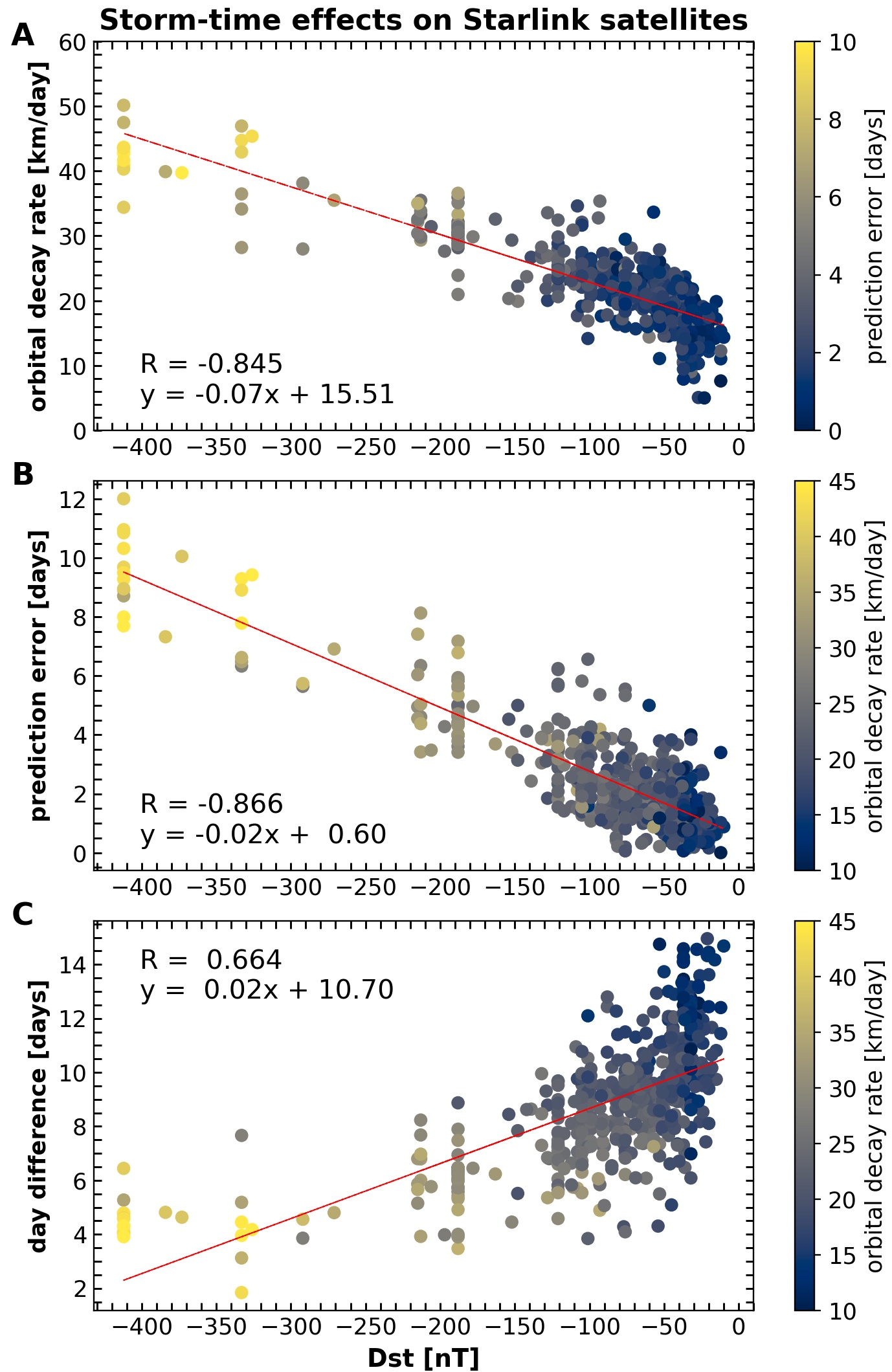}
                        \caption{Scatter plots of geomagnetic activity effects on 523 Starlink reentries in the period 2020-2024. {\bf (A)}: orbital decay rate as a function of Dst (colors indicate prediction errors); {\bf (B)}: prediction error as a function of Dst (colors indicate orbital decay rate); an {\bf (C)}: day difference as a function of Dst (colors indicate orbital decay rate). Prediction error is the difference between the reentry and predicted epochs, whereas day difference is the difference between the reentry and reference altitude epochs. The $R$ values shown in the plot represent Pearson's correlation coefficients.}
                        \label{sl_scatter}
                \end{figure}

            Statistical results of Starlink reentry parameters occurring during different geomagnetic activity levels are documented in Figure \ref{sl_scatter}. Panel A shows orbital decay rate as a function of Dst color coded by prediction errors. Panel B shows prediction errors plotted as a function of Dst color coded by orbital decay rate. Finally, panel C shows day difference plotted as a function of Dst color coded by orbital decay rate. \par

            Although there are limited reentries under Dst $<$ --150 nT conditions, it is very clear from all frames that there is a strong correlation between orbital decay rate with Dst (A), and prediction error with Dst (B). The colors in (A) represent the variable plotted in (B) and vice versa, thus both panels complement each other. This explains the high Pearson correlation coefficients for panels A and B, $R$ = --0.85, and $R$ = -0.87, respectively. However, frame C shows a weaker correlation ($R$ = 0.66) between day difference and Dst. This can be explained by the large error bars shown in all panels in Figure \ref{xht_sea}, particularly during the reentries ($t$ $>$ 0).  \par

            \begin{figure}
                \centering
                \includegraphics[width = 11cm]{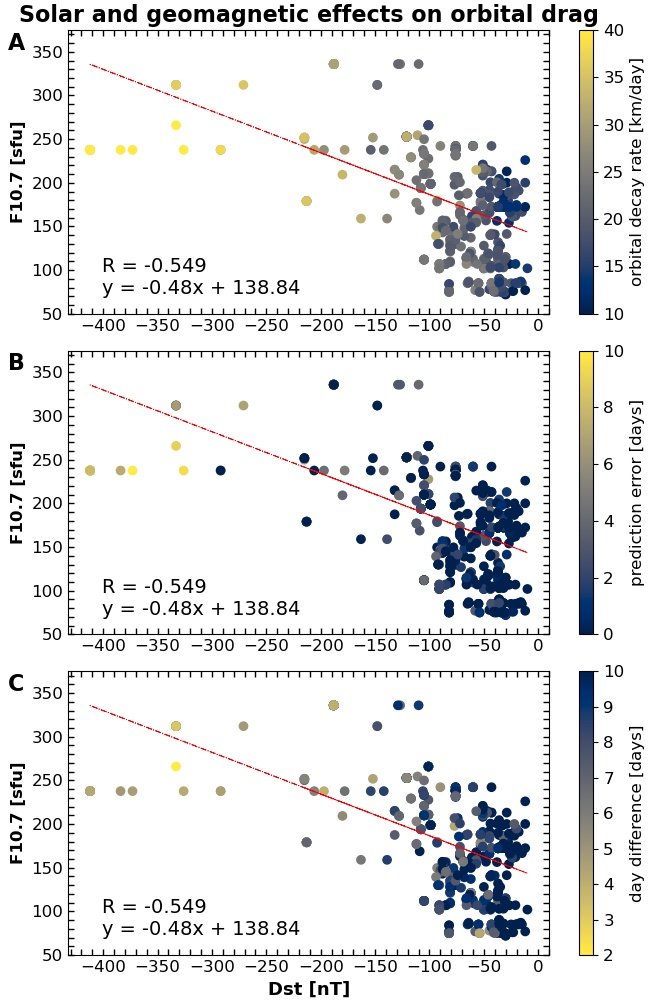}
                \caption{Effects of solar radiation and geomagnetic activity on the reentries of the 532 Starlink satellites investigated in this study. All panels show Dst plotted as a function of F10.7, with the colors representing {\bf (A)}: }
                \label{dst_f10}
            \end{figure}

            Despite having many reentries clustered with Dst $>$ --150 nT, a reasonable linear relationship can be seen for orbital decay rate as a function of Dst. These linear relationships are found in each panel. As a result, average orbital decay rates can be predicted as a function of Dst (panel A). A similar linear relationship can be used for average prediction errors during reentries (panel B). Although the correlation coefficient for day difference is the lowest, the linear relationship shown in panel C can be used for coarse predictions of day differences between estimated reentry epoch and reference altitude epoch. However, such linear relationships can be further improved by using Starlink ephemeris data with much finer resolution than a single day for TLEs. \par

            Figure \ref{dst_f10} shows effects caused by solar activity in combination to geomagnetic activity on Starlink orbital drag. Solar activity is indicated by F10.7 solar radio flux index data, whereas geomagnetic activity is indicated by Dst index data. This is plotted in all panels, but the color codes indicate: orbital decay rate {\bf (A)}; prediction error {\bf (B)}; and day difference {\bf (C)}. The Dst value associated with each reentry is the same as explained before, but the F10.7 values associated with each reentry corresponds to the maximum value in a time interval of 14 days around the reference altitude epoch. We chose this time window to accommodate 27-day secular trends of solar radiation on thermospheric neutral mass density. \cite{Emmert2015} provides many references of empirical thermospheric models that use this long-term secular trends in thermospheric density determinations. \par

            Results show a trend of increasing geomagnetic activity with solar activity, but the correlation coefficient between Dst and F10.7 is relatively low ($\sim$ 0.55). This can be explained by the fact that halo (wide) CMEs can also occur during high solar activity of a relatively weak solar cycle, such as in the case of current SC25 \citep{Gopalswamy2023}. Halo CMEs are generally less geoeffective than those bound more directly towards Earth because their broader geometry means they are less likely to hit Earth's magnetosphere with the same intensity, hence triggering weaker geomagnetic storms \citep{Besliu-Ionescu2021}. Despite the moderate correlation depicted in Figure \ref{dst_f10}, it is clear from the figure that high orbital decay rates and prediction errors, along with low day differences, usually occur during high solar activity periods. A direct correlation analysis between F10.7 and the orbital drag parameters shown in Figures \ref{xht_sea} is not shown here because the resulting correlation coefficients are too low ($\sim$ 0.25). This may be explained by the fact that, during geomagnetic storms, Joule heating significantly surpass solar radiation as energy source into the ionosphere-thermosphere system \citep{Knipp2004,Prolss2011,Emmert2015}. Therefore, the former has significantly larger impacts on short-term satellite orbital drag effects in comparison to the latter. \par

            Another possible effect that may have impacted our results, particularly in the occurrence of large altitude error bars and the scatter of day difference (Figure \ref{sl_scatter}C), comes from variations of the areas and masses of Starlink satellites. The ballistic coefficient of a satellite is a measure of its ability to resist atmospheric drag, with a higher coefficient indicating a more streamlined satellite that experiences less resistance during its orbit \citep{Sutton2005,Oliveira2019b,Bruinsma2023}. The ballistic coefficient is directly proportional to the drag coefficient and the area-to-mass ratio. The solar panel areas and masses of Starlink satellites have increased over time to support more advanced technologies and improve satellite performance for enhanced service quality, including faster speeds and greater capacity for users. The masses and areas of Starlink satellites have increased by a factor of four since its first version (\url{https://www.space.com/spacex-starlink-satellites.html#section-spacex-s-plans-for-starlink}). As a result, different drag effects would occur on satellites with different ballistic coefficients in regions with similar thermosphere neutral mass density, hence the satellites' altitudes would also be different. For this reason, a superposed epoch analysis using Starlink satellites with different ballistic coefficients would enhance the quality of our results. \par

            \add{\protect{The duration of geomagnetic storms can also play a role in determining the severity of drag effects on LEO satellites. For example, \cite{Oliveira2020b} demonstrated that storm duration can play an equally or more critical role than intensity in driving drag effects. By analyzing historical superstorms, such as those in March 1989 and December 1921, the authors found that despite being less intense, the longer-lasting March 1989 storm caused orbital decays up to 400\% greater than the shorter-lasting 1921 event. This underscores that prolonged exposure to elevated thermospheric densities during extended storms can lead to more severe satellite drag than shorter, more intense storms. Such interplay between storm intensities and durations may also have played a role in our results, particularly in the larger altitude error bars and day difference scattering mentioned above. However, a direct quantitative analysis of storm duration impacts on our results is difficult to be achieved with our data sets, particularly in the way the data are organized. For example, in their simulation, \cite{Oliveira2020b} lined up the simulated thermospheric mass density (and drag effects) with the SSC of each event, in a similar way as shown in Figure \ref{starlink_may2024} and Figure 2 of \cite{Oliveira2025a}. The reference altitude epoch and SSC/storm main phase epoch do not necessarily align for a specific reentry, and our events would have to be inspected on a case-by-case basis. Additionally, some of the reentries here investigated may have occurred during geomagnetic activity driven by corotating interaction regions (CIRs). This may add another layer of complexity to the problem, since timing the beginning and end of a CIR can also be tricky \citep{Gosling1999}. Discriminating reentries during CME- and CIR-driven geomagnetic activity is also important because CME-driven storms are generally more intense and qre briefer than CIR-driven storms \citep{Borovsky2006}.}}

            As a result, in general, the statistical and superposed epoch analysis results shown in Figures \ref{xht_sea}-\ref{dst_f10} agree with the case sample shown in Figure \ref{starlink_may2024} and with the comparative case study of \cite{Oliveira2025a}. Therefore, our results can be used for predictions of Starlink satellite reentry epochs as a function of geomagnetic activity. \par

\section{Conclusion}

    In this work, we derived altitudes and velocities from TLE data to explore geomagnetic activity effects on reentries of Starlink satellites from VLEO in the period 2020 to 2024. By using Dst data to represent geomagnetic activity, we investigated storm effects on the orbits of 523 satellites as a function of storm intensity and solar activity. By taking a reference altitude between 260-320 km for each reentry, we superposed altitude and velocity data 30 and 16 days around the zero epoch time. We also compared orbital decay rates, day prediction errors between reentry prediction epoch and estimated reentry epoch, and day differences between estimated reentry epochs and reference altitude epoch as a function of Dst. Our main findings are listed below:
    
    \begin{enumerate}

        \item As pointed out by \cite{Oliveira2021a}, we observed a large number of satellites reentering from VLEO during high level of solar activity in the rising phase of SC25. Such combined observations are unprecedented in the history of human activity in LEO. 

        \item Geomagnetic storms can directly affect how fast a satellite reenters. The higher the geomagnetic activity level, the faster the satellite reenters. This is clearly shown in Figure \ref{xht_sea} for the day difference between the altitude reference and the K\'arm\'an line crossing epochs. This is explained by increasing orbital decay rates as a function of geomagnetic activity. Our statistical results confirm the case study provided by \cite{Oliveira2025a}.

        \item The day difference between reentry prediction epochs (derived with the SGP4 orbit propagator) and estimated reentry epochs from TLE data also increases with geomagnetic activity (prediction error). This finding reinforces the importance of using accurate orbital drag models for collision avoidance in LEO and VLEO and safeguard the ground from satellite debris impacts generated from their reentries.

        \item The velocity results shown in Figure \ref{vel_sea} confirm the findings described before. The higher the geomagnetic activity, the faster the satellite reaches the average K\'arm\'an line velocity (7.85 km/s).

        \item The findings summarized in items 2 and 3 are confirmed by the scatter analyses shown in Figure \ref{sl_scatter}. This is explained by the fact that faster orbital decay rates are associated with large predicton errors (how sooner a satellite reenters with respect to prediction) and day difference (how sooner a satellite reenters with respect to reference altitude epoch). Our results can be used to predict the reentry epochs of Starlink satellites as a function of geomagnetic activity levels \add{from a 11last chance", reference altitude near 280 km}.

        \item Solar activity also impacts orbital drag effects during the Starlink reentries. Although the correlation between F10.7 and Dst is moderate, the trends of higher orbital drag effects (orbital decay rates, prediction error, and day difference) are seen for higher F10.7 observations (Figure \ref{dst_f10}).

    \end{enumerate}

    The effects caused by geomagnetic storms on thermosphere neutral mass density and satellite orbital drag \citep{Jacchia1959,Prolss2011,Emmert2015} and on their predictions (including reentries) have been known for decades \citep{Doornbos2006,Klinkrad2006,Choi2017,Geul2018,He2018,Oliveira2021a}. All investigations hitherto have presented case studies or superposed epoch analysis studies involving one or two satellites \citep{Sutton2005,Doornbos2012,Oliveira2019b,Krauss2020,Bruinsma2021}. However, our study is the first to provide strong evidence of geomagnetic activity effects on orbital drag and subsequent reentries of more than 500 similar satellites, specifically SpaceX’s Starlink spacecraft. This includes many satellites reentering during similar times. Therefore, even though using limited daily TLE data, we show that satellites decay faster as the storm becomes more intense. Our results are also supported by the study of \cite{Parker2024}, who observed with limited daily TLE data that nearly half of 10,000 payloads in LEO (mostly Starlink objects), performed a large-scale {\it en masse} maneuver during the May 2024 event, the largest satellite migration in history. Our results are promising because they point in the direction of using short-cadence Starlink data (precise orbit determination, neutral mass density, ram direction area, drag coefficient) for the improvement of orbital drag models during geomagnetic storms, particularly during extreme events \citep{Oliveira2021a}.
    
\section*{Data Availability Statement}
    
    The Dst data, provided by \cite{WDC_Dst2015}, was obtained from \url{https://wdc.kugi.kyoto-u.ac.jp/dst_realtime/index.html}. The Starlink TLE data for altitudes and velocities was downloaded from \url{space-track.org} by clicking on the tab ``ELSET Search". The F10.7 solar radio flux data, provided by a collaboration between the National Research Council Canada and the Natural Resources Canada, was downloaded from \url{https://www.spaceweather.gc.ca/forecast-prevision/solar-solaire/solarflux/sx-5-en.php}.

\section*{Conflict of Interest Statement}

    The authors declare that the research was conducted in the absence of any commercial or financial relationships that could be construed as a potential conflict of interest.

\section*{Author Contributions}

    This original research article was written by the first and second authors. The final version of this manuscript was read and approved by all the authors.

\section*{Funding}
    
    DMO and EZ acknowledge financial support provided by NASA's Space Weather Science Applications Operations 2 Research. DMO thanks UMBC for providing financial support through the START (Strategic Awards for Research Transitions) program (grant \# SR25OLIV).


\end{document}